\pdfoutput=1
\documentclass{PoS}

\usepackage{amssymb}
\usepackage{amsmath}
\newcommand{\Lagr}{\mathcal{L}}
\newcommand{\D}{\mathrm{d}}

\title{Standard Model vacuum stability in the presence of gauge invariant nonrenormalizable operators} 

\ShortTitle{SM vacuum stability with gauge invariant nonrenormalizable operators}

\author{Zygmunt Lalak\\
        Institute of Theoretical Physics, Faculty of Physics, University of Warsaw ul. Pasteura 5, 02-093 Warsaw, Poland\\ 
        E-mail: \email{Zygmunt.Lalak@fuw.edu.pl}}

\author{\speaker{Marek Lewicki}\\
        Institute of Theoretical Physics, Faculty of Physics, University of Warsaw ul. Pasteura 5, 02-093 Warsaw, Poland\\
        E-mail: \email{Marek.Lewicki@fuw.edu.pl}}
        
\author{Pawe\l{} Olszewski\\
        Institute of Theoretical Physics, Faculty of Physics, University of Warsaw ul. Pasteura 5, 02-093 Warsaw, Poland\\
        E-mail: \email{Pawel.Olszewski@fuw.edu.pl}}

\abstract{
Analysis of the Standard Model effective potential can reveal certain properties of the UV completion of SM. We discuss the issue of stability of the SM potential after inclusion of new nonrenormalizable interactions. We present the map of lifetimes resulting from different configurations of the new operators. The key result is proof of gauge independence of our results after inclusion of nonrenormalizable operators. We also discuss the consequences of RGE improvement of the potential.

}

\FullConference{18th International Conference From the Planck Scale to the Electroweak Scale \\
		25-29 May 2015\\
		Ioannina, Greece }

\begin{document}
\section{Introduction}
The first run of LHC was a great success due to the discovery of Higgs Boson and measurement of its properties which finally complete our knowledge of the SM effective potential.
Paired with lack of other discoveries at the LHC, investigation of the structure oft the potential becomes one of the most promising new opportunities towards new physics searches.
This line of research already received significant attention  \cite{Coleman,Buttazzo:2013uya,Degrassi,Ellis:2009tp,Espinosa,Casas:2000mn,Casas:1996aq,Casas:1994qy,Sher,Branchina:2013jra}. 

The study of the renormalization group improved effective SM  potential has revealed an interesting structure at field strengths  higher than approximately $10^{11}$ GeV and new minima at superplanckian field strengths. The upshot depends critically on the precise value of the measured Higgs mass and on the measured value of the top quark Yukawa coupling. 
In particular, one finds that for the central value of the top mass and for the central value of the measured Higgs mass the physical electroweak symmetry breaking minimum becomes metastable with respect to the tunneling from the physical EWSB minimum to a deeper minimum located at superplanckian values of the Higgs field strength. The lifetime of the metastable SM Universe is much larger than the presently estimated age of the Universe, however  the instability border in the space of parameters $M_{top} - M_{higgs}$ looks uncomfortably close, which suggests that the result is rather sensitive to modifications that can be brought in by the BSM extensions. 

The question of stability in the presence of nonrenormalizable operators was raised in \cite{Branchina:2013jra}.
This question has been studied further in \cite{Lalak:2014qua}. In particular a map of the vacua lifetime in SM with new nonrenormalizable scalar couplings was presented.
Also RGE improvement of the new couplings, have been included, and shown to have a significant impact on the resulting lifetime.
The results were also obtained by a fully numerical approach instead of the analytical approximations previously used in the literature. 

Here we continue the discussion of gauge dependence of the vacuums lifetime \cite{Lalak:2015usa,Espinosa:2015qea}.
In particular we show that determination of the tunnelling rate remains gauge independent after inclusion of new nonrenormalizable operators.
\section{Gauge independence of the vacuum lifetime}

The effective potential, and in general the whole effective action of a gauge field theory, is a gauge-dependent object. Attempting a computation, one finds himself at liberty to choose both the form of a gauge fixing Lagrangian and numeric values of parameters therein.  Nevertheless, just like using gauge-dependent Green functions one arrives at unambiguous scattering amplitudes, proper usage of the effective potential yields observables independent of the gauge fixing. This statement applies also in the framework of any perturbative expansion, where absence of gauge-dependent terms means only that they are of higher order than the employed accuracy.

One example of such observables inferred from the potential, especially relevant in the context of radiative symmetry breaking and phase transitions, are its values at the extrema. But positions of the extrema, as well as any other values of the field, are gauge-dependent. Another quantity one computes using the effective Lagrangian, is the tunneling rate between local minima of the potential via instanton known as Coleman bounce \cite{Coleman}. Although there are results that support the belief that this rate is gauge-independent \cite{Metaxas:1995ab}, no comprehensive proof of that exists to our knowledge. Moreover the gauge-independence seems to be often hopelessly lost on a formal level in the course of approximations used in practice. Possibly the most simplified estimate for the lifetime of electroweak vacuum relative to the age of the universe $T_U$ is
\begin{equation}\label{lifetime}
\frac{\tau}{T_U} \sim \frac{1}{\Lambda^4 T_U^4} e^{\frac{8 \pi^2}{3}\frac{1}{|Z^2(\Lambda) \lambda(\Lambda) |} }\;,
\end{equation}
where
\begin{equation}
Z(\mu)^{\frac{1}{2}} = e^{-\int \gamma(\mu) \,\D \log(\mu)}
\end{equation}
denotes running of the field via a nonzero anomalous dimension, and $\lambda(\mu)$ is the running coupling. The scale $\Lambda$ is chosen so as to minimize the negative value of $ \lambda(\Lambda)$. The rationale behind this formula is as follows. The classical Lagrangian
\begin{equation} \label{classlagr}
\Lagr = \frac{1}{2}\left( \partial \phi \right)^2 - \frac{\lambda_c}{4} \phi^4 \;,
\end{equation}
where this time $\lambda_c$ is a negative constant, admits a bounce solution governing decay of the $\phi = 0$ configuration, whose action is $S=\frac{8 \pi^2}{3}\frac{1}{\lambda_c}$. Hence, after reinterpretation of the Lagrangian \eqref{classlagr} as an quantum effective Lagrangian, the sketchy formula
\begin{equation}\label{tunnelrate}
\rho = (\text{dimensionfull quantity})^4 \,e^{-\frac{8 \pi^2}{3}\frac{1}{|\lambda_c|}}
\end{equation}
gives space-density of the tunneling rate. Due to classical scale invariance of the Lagrangian, one has no dimensionfull quantity to fill in above. In a more honest approach the effective Lagrangian would be rather
\begin{equation} \label{quantlagr}
\mathcal{L} = \frac{1}{2}Z(\mu)\left( \partial \phi \right)^2 - \frac{\lambda(\mu)}{4} Z(\mu)^2 \phi^4 \;.
\end{equation}
This form is correct only at the lowest level of perturbative calculation, when logarithmic loop correnctions are not yet included. Going to higher loop orders introduces more complicated dependence on $\phi$, as well as explicit presence of the renormalization scale $\mu$. Now we notice that, for one, the full (both running and loop corrected) coefficients in front of the kinetic and quartic terms are dimensionless and so, absent any dimensionfull parameters in the theory, they may depend on $\mu$ only via $\phi/\mu$. Secondly the running of $Z$ and $Z^2 \lambda$ fully captures these coefficients' dependence on $\mu$, in the sense that it cancels between running couplings and explicit dependence of loop corrections. In conclusion, it is justified to throw away $\mu$ from \eqref{quantlagr}, replace it with the field, $\mu \rightarrow \phi$, and claim that the resulting $\mu$-independent function of $\phi$ well approximates the full effective quantum Lagrangian (or at least part of it containing up to two field derivatives). This leaves us with
\begin{equation} \label{quantlagr2}
\Lagr = \frac{1}{2}\left( \partial Z^{\frac{1}{2}}(\phi) \phi \right)^2 - \frac{\lambda(\phi)}{4} Z^2(\phi) \phi^4 \;.
\end{equation}

Mimicking \eqref{tunnelrate}, one may choose to take as the tunneling rate, the expression
\begin{equation} \label{tunnelrate2}
\rho \sim {\Lambda^4 } e^{-\frac{8 \pi^2}{3}\frac{1}{|Z^2(\Lambda) \lambda(\Lambda) |} }\;.
\end{equation}
It is the quantum corrections that break the scale invariance of the classical Lagrangian, effectively pushing the counterpart of $\lambda_c$ toward positive values. Taking the minimum of $\lambda(\mu)$ both minimizes the exponential factor and approximately marks energy scale $\Lambda$, where the potential starts becoming convex again. The fact, that this choice turns out to be exceptionally good, is more formally investigated in \cite{DiLuzio:2015iua}.
Finally, approximating the four-volume of our past lightcone by age of the universe raised to the forth power, and taking an inverse to switch from decay rate to lifetime, we arrive at \eqref{lifetime}. Error introduced by not caring about order one factors in front of the exponential function typically pales in comparison with uncertainty of the exponent.

This brings us to our point. $|Z^2 \lambda|$ in  \eqref{lifetime} should be replaced by $|\lambda|$ alone.
\begin{equation}\label{lifetime2}
\frac{\tau}{T_U} \sim \frac{1}{\Lambda^4 T_U^4} e^{\frac{8 \pi^2}{3}\frac{1}{| \lambda(\Lambda) |} }\;,
\end{equation}
 From the point of view of the Lagrangian \eqref{quantlagr2}, it makes more sense to redefine the field variable by $\tilde{\phi} = Z^{\frac{1}{2}}(\phi) \phi$, thus completely eliminating $Z(\mu)$. But this has an even more appealing advantage on its own. As is known, among running parameters, vertex-like couplings, including $\lambda(\mu)$, do not depend on gauge fixing. Conversely, the anomalous dimension is gauge-dependent, thus making \eqref{tunnelrate2} very sensitive to the values of gauge fixing parameters. The absence of $Z$ in \eqref{lifetime2} trivially makes it gauge-independent.
 
 This approach is in line with the treatment of abelian gauge theory discussed in \cite{Metaxas:1995ab}, where it was showed that, as far as formal consistency is concerned, including quantum corrections to the kinetic term is crucial. The reason is that the gauge dependence of instanton's action cancels between contributions from potential and the kinetic term. Putting $Z$ in the exponent of \eqref{lifetime} is an omission resulting from approximating the effective Lagrangian by sum of classical kinetic term and quantum potential.

Straightforward as it sounds, the idea to throw away $Z$'s may seem controversial, because it advertises treating $\mu$ in $Z(\mu)$, substituted by $\phi$, as a spacetime dependent configuration, and thus hitting $Z(\phi(x))$ with $\partial_\mu$.

\begin{figure}[ht]
\centering
\includegraphics[scale=1.0]{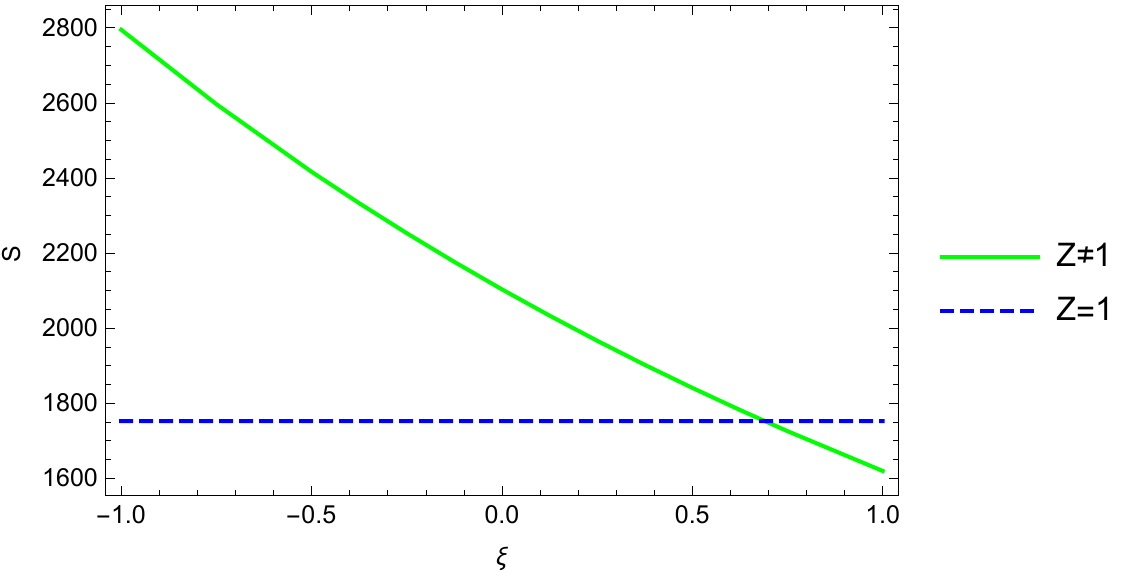} 
\caption{Solid curve: gauge dependence of the bounce action calculated without $Z$ factor in front of the kinetic term. Dashed line: the bounce action calculated after the redefinition $\tilde{\phi} = Z^\frac{1}{2}(\phi) \phi$. $\xi$ is a gauge fixing parameter, see text.} 
\label{SodXi}
\end{figure}

In the case of Standard Model, difference between \eqref{lifetime} and \eqref{lifetime2} is significant. $Z^2(\mu)$ drops down from $ 1$ at the scale of the top mass, $M_{top}$, to about $0.8$ closer to the Planck mass, when computed in Landau gauge. This translates to the exponent in \eqref{lifetime} (the bounce's action) increasing from roughly $1800$ to $2100$. Correspondingly the relative vacuum's lifetime computed via \eqref{lifetime2} is only around $10^{529}$ as compared to $10^{676}$ when using \eqref{lifetime}. Additional fig. \ref{SodXi} illustrates gauge dependence of the action, for the simplest class of gauge fixing Lagrangian, so called Fermi gauges, with $\xi = \xi_W(M_{top})=\xi_B(M_{top})$, (see e.g. \cite{DiLuzio:2014bua}). Landau gauge belongs to this this class as an RGE-stable choice of $\xi =0$.

Now consider including nonrenormalizable terms to the scalar potential, e.g. 
$
\frac{\lambda_6}{6!\, M^2} \phi^6\;.
$
In our temporarily simplistic approach to quantum corrections, the effective contribution of this term reads
\begin{equation}
\frac{\lambda_6(\phi)}{6!\, M^2} Z^3(\phi) \phi^6\;,
\end{equation}
and $\lambda_6$ contributes to RGE's of other couplings. But obviously it does not appear in $\gamma$ at one loop. Since, again, running of $\lambda_6(\mu)$ does not depend on gauge, after absorbing $Z$'s into redefinition of the field, no gauge dependence is left in this term.

There is yet more sophisticated point to be made about gauge dependence of nonrenormalizable terms in the potential. Recall that Nielsen identities bind variation of the effective action with its derivative wrt. gauge parameters \cite{Nielsen:1975fs}. Taking momentumless part of this relation (as is usually done when going from effective action to effective potential), one arrives at
\begin{equation}\label{Nielsen}
\xi \frac{\partial V(\phi)}{\partial \xi} = C^\xi (\phi) \frac{\partial V(\phi)}{\partial \phi}\;.
\end{equation}
The $C$ function is not guaranteed not to blow up for some unlucky choice of gauge fixing, but let us assume that does not happen.
The point is that $C$ admits a perturbative expansion, just like other parts of the equation \eqref{Nielsen}, but then it inherently involves ghost, goldstone and gauge propagators \cite{Aitchison:1983ns}. Consequently vertices produced by nonrenormalizable operators, like six-legged $\lambda_6$, do not contribute to $C$ at lowest levels of the expansion. When we now classify terms at the left and right hand side of \eqref{Nielsen} by their power of $M$, we see that the very same $C$ function separately governs gauge dependence of both renormalizable and nonrenormalizable part of the potential of any dimension. In conclusion, for practical purposes, it is enough to uncover gauge dependence of the renormalizable part, to fully reconstruct it for operators of all dimensions.
\section{New interactions and Standard Model  phase diagram}
We will consider the SM Lagrangian with two new nonrenormalizable operators $|H|^6$ and $|H|^8$, where $H$ is the Higgs doublet. Both operators are suppressed by a mass scale $M$ to an appropriate power.
We constrain our analysis to the real Higgs field $H=(\phi / \sqrt{2},0)$, which results in the following potential
\begin{equation}\label{potential}
V=-\frac{m^2}{2} \phi^2 +\frac{\lambda}{4} \phi^4 + \frac{\lambda_6}{6!}\frac{\phi^6}{ M^2} + \frac{\lambda_8}{8!}  \frac{\phi^8}{M^4}.
\end{equation}

\begin{figure}[ht]
\begin{minipage}[t]{0.29\linewidth} 
\centering
\includegraphics[scale=0.8]{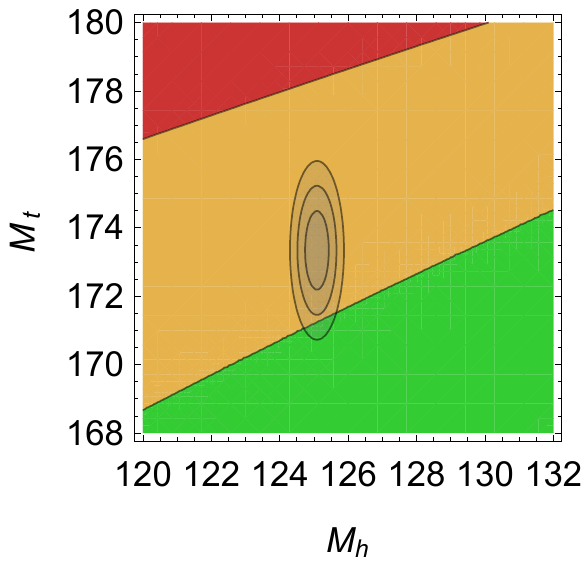} 
\end{minipage}
\hspace{0.5cm}
\begin{minipage}[t]{0.33\linewidth}
\centering 
\includegraphics[scale=0.8]{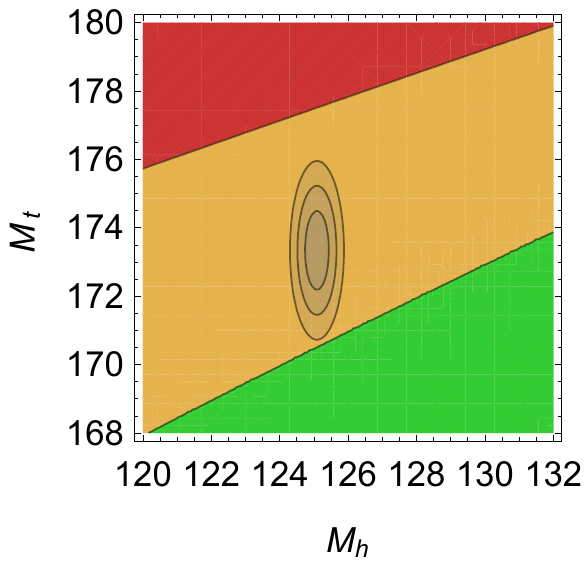}  
\end{minipage}
\begin{minipage}[t]{0.31\linewidth}
\centering 
\includegraphics[scale=0.8]{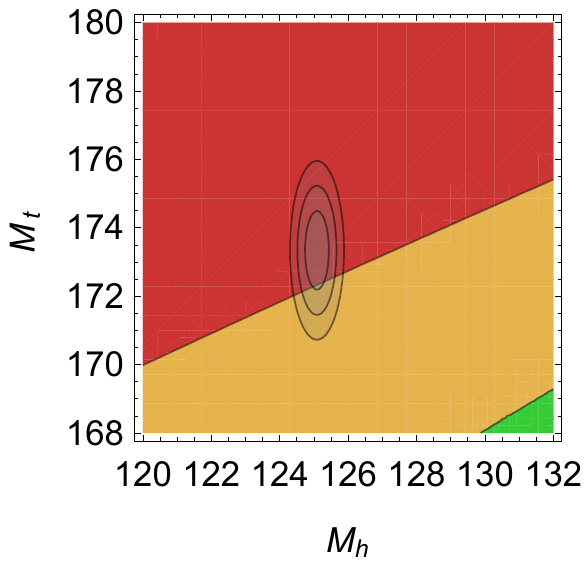}  
\end{minipage}
\caption{
SM phase diagram (left panel), SM phase diagram including new operators suppressed by the Planck mass, $\lambda_6(M_p)=-1/2$ and  $\lambda_8(M_p)=1$ (middle panel) and $\lambda_6(M_p)=-1$ and  $\lambda_8(M_p)=1/2$ (right panel). The red region corresponds to instability, the green region to absolute stability, and the yellow region is metastable.
\label{phasediagram}
}
\end{figure}

It is well known that radiative corrections to SM couplings have a crucial effect on the vacuum stability, and they require precise determination \cite{Buttazzo:2013uya}.
It is therefore important to determine the effects of radiative corrections on the new couplings.
Their One-loop beta functions take the form
\begin{eqnarray} \label{betafunctions}
16\pi^2\beta_{\lambda_6} &=& \frac{10}{7}\lambda_8\frac{m^2}{M^2}+18\lambda_6 6\lambda-6\lambda_6\left(\frac{9}{4}g_2^2+\frac{9}{20}g_1^2-3y_t^2\right), \\ 
16\pi^2\beta_{\lambda_8} &=& \frac{7}{5}28\lambda_6^2+30\lambda_8 6\lambda-8\lambda_8\left(\frac{9}{4}g_2^2+\frac{9}{20}g_1^2-3y_t^2\right), \nonumber
\end{eqnarray}
which agrees with \cite{Jenkins:2013zja}.
Figure~\ref{phasediagram} illustrates effects of new nonrenormalizable operators on Standard Model vacuum stability in the form of the well known SM phase diagram (see for example \cite{Buttazzo:2013uya}) and the same diagram influenced by certain values of the new operators, respectively  $\lambda_6(M_p)=-1/2, -1$ and  $\lambda_8(M_p)=1, 1/2$.
Above plots clearly show that nonrenormalizable interactions suppressed  by the Planck mass can dramatically change the SM phase diagram. If they induce a new deep minimum around the Planck scale the electroweak vacuum is pushed towards the instability region. 

\section{Lowering the magnitude of the suppression scale}
\begin{figure}[ht]
\begin{minipage}[t]{0.47\linewidth}
\centering
\includegraphics[scale=0.6]{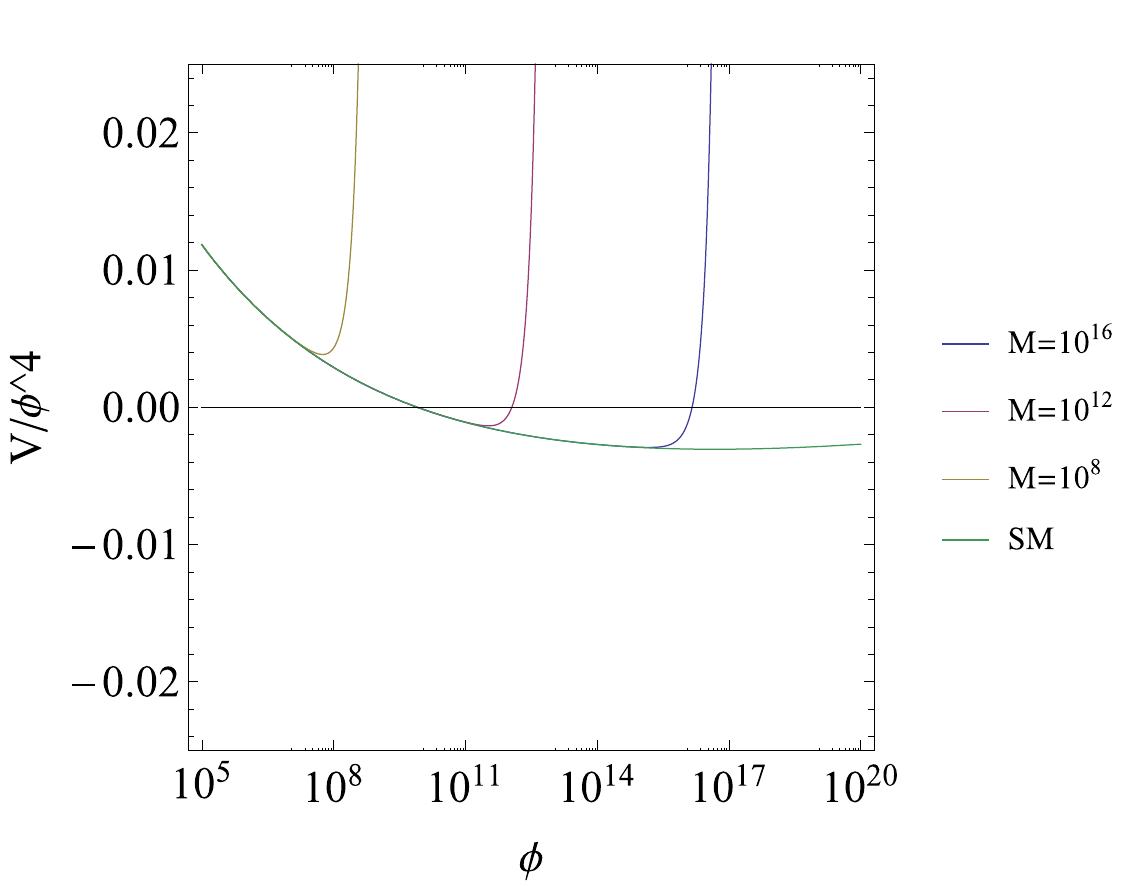}
\end{minipage} 
\begin{minipage}[t]{0.47\linewidth}
\centering
\includegraphics[scale=0.6]{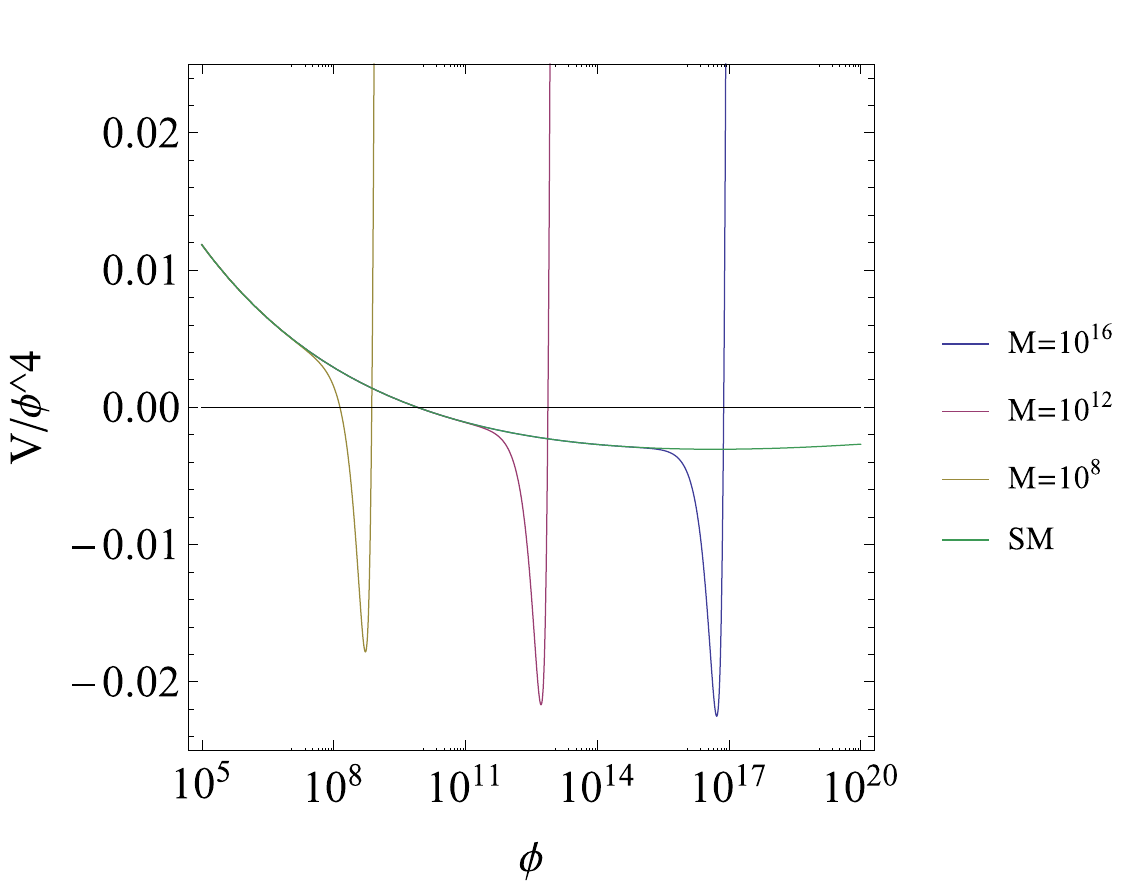}
\end{minipage} 
\caption{Scale dependence of $\lambda_{eff}/4 = V/\phi^4$ with $\lambda_6=\lambda_8=1$ (left panel) and $-\lambda_6=\lambda_8=1$ (right panel) for different values of suppression scale $M$. The lifetimes corresponding to suppression scales $M=10^{8},10^{12},10^{16}$ are respectively, $\log_{10}(\frac{\tau}{T_U})=\infty,1302,581$ (left panel) and $\log_{10}(\frac{\tau}{T_U})=-45,-90,-110$ (right panel) while for the Standard Model $\log_{10}(\frac{\tau}{T_U})=540$.
\label{scaleplots1}
}
\end{figure}

It is interesting to see how our results would change for lower suppression scale $M$ of the new operators in (\ref{potential}) .  To perform a qualitative analysis it is sufficient to use the analytical approximation (\ref{lifetime2}). When both nonrenormalizable operators are positive, lowering the suppression scale $M$ corresponds simply to lifting the potential not far above $M$. Then the action (the exponent in (\ref{lifetime2})) increases. This happens because the position of the minimum of $\lambda_{eff}$ shifts towards smaller energy scales decreasing the value of $|\lambda_{eff}|$, as we can see in left hand side panel of Figure~\ref{scaleplots1}. 
 
When nonrenormalizable operators induce a new deeper minimum, which corresponds to positive $\lambda_8$ and negative $\lambda_6$ this dependence is smaller.
Now changing the scale, changes $\lambda_{eff}$ by a small fraction of its value which means the resulting lifetimes are much less scale dependent, as shown in the right hand side panel of Figure~\ref{scaleplots1}. 
 
The last possibility is a potential unbounded from below ($\lambda_8<0$) which corresponds to unacceptable unstable solutions (see \cite{Lalak:2014qua}).

\section{Summary}
In this note we have discussed the issue of gauge dependence of expected lifetime of the electroweak vacuum. We have shown the method of obtaining a gauge invariant result even after inclusion of new nonrenormalizable operators.
We have presented a map of the lifetime of vacua in the SM extended by such operators, taking into account their running.
We have also performed a full numerical calculation going beyond the standard analytical approximations used when calculating the lifetime of the metastable vacuum.

We have shown that it is easy to destabilise the SM vacuum with respect to new minimum appearing at the Planck scale due to new scalar operators. 
While taking into account running of the new couplings slightly ameliorates this problem, the generic effect is still shortening of  the lifetime of the electroweak vacuum. Our results also show the dependence of the lifetime on the magnitude of the suppression scale of new couplings.
In particular we have demonstrated that absolute stability can be achieved only by lowering the suppression scale of nonrenormalizabe operators while picking combinations of new couplings, which do not generate new deep minima in the potential 
\begin{center}
{\bf Acknowledgements}
\end{center}
This work was supported by National Science Centre 
under  research grants DEC-2012/04/A/ST2/00099 and  DEC-2014/13/N/ST2/02712.
ML was supported by the Polish
National Science Centre under doctoral scholarship number
2015/16/T/ST2/00527

\end{document}